\ifdefined\pdfminorversion
\fi
\documentclass[aps,prb,reprint,amsmath,amssymb,floatfix,longbibliography]{revtex4-2}
\usepackage{array}[=2016-10-06]
\usepackage{graphicx}
\usepackage{bm}
\usepackage{booktabs}
\usepackage{dcolumn}
\usepackage{xcolor}
\usepackage[colorlinks=true,citecolor=blue,linkcolor=blue,urlcolor=blue,filecolor=blue,hypertexnames=false]{hyperref}
\newcommand{\smsec}[1]{\ref{sec:#1}}
\hypersetup{pdftitle={Equal-energy magnons with unequal lifetimes: Orbital-pair scattering and a comparison with amplitude-mode decay in Ca2RuO4},pdfauthor={Jyotirmoy Sau}}
\newcommand{\Rea}{\operatorname{Re}}
\newcommand{\Ima}{\operatorname{Im}}
\newcommand{\fig}[3]{%
\begin{figure*}[t]
\centering\includegraphics[width=#2\textwidth]{figures/fig#1.pdf}
\caption{#3}\label{fig:#1}
\end{figure*}}
\begin{document}
\title{Equal-energy magnons with unequal lifetimes:
Orbital-pair scattering and a comparison with amplitude-mode decay in Ca$_2$RuO$_4$}
\author{Jyotirmoy Sau}
\email{jyotirmoy.sau@physics.uu.se}
\affiliation{Department of Physics and Astronomy, Uppsala University, Box 516, SE-75120 Uppsala, Sweden}
\date{November 25, 2026}
\begin{abstract}
Two magnons can carry opposite spin and have the same resonance energy, yet one can lose intensity much faster than the other. In the square-lattice model studied here, a magnon can emit two orbital excitations and continue at lower energy with its spin intact. The probability of emission depends on the direction of motion and on the magnon spin. An adjustable exchange anisotropy brings the two resonance energies together while their linewidths remain different. A weak-coupling calculation identifies the origin of this difference. At a representative finite coupling, the ratio of the linewidths is $14.30$ when at most one orbital pair is present at a time. Including all states with up to two pairs gives an extrapolated ratio of $15.53$ as the numerical broadening tends to zero. For flat orbital modes in the one-pair calculation, the energies coincide along a contour in momentum space, and a quarter turn interchanges the longer-lived spin branch. An initially unpolarized beam can therefore develop a spin imbalance among the magnons that remain in its original energy window. A separate calculation for Ca$_2$RuO$_4$, where an amplitude mode couples to two transverse magnetic modes, illustrates how the energies of the final states shape a measured spectrum.
\end{abstract}
\maketitle
\raggedbottom

\section{Introduction}

A magnon is a quantum of a spin wave, a collective oscillation of the spins in a magnet. In an energy-resolved spectrum, its resonance energy locates the excitation, while its linewidth tells us how quickly an initially prepared mode loses intensity. Interactions affect both quantities. For example, coupling to lower-energy excitations can shift a magnon resonance and give it a finite lifetime when decay is allowed~\cite{cz2009}. Consider two magnons carrying opposite spin at the same momentum. If their resonance energies are tuned to coincide, can their lifetimes still differ?

Electronic orbitals provide a way to address this question. Their shapes determine how neighboring magnetic ions exchange spin, so an ordered arrangement of orbitals can make the exchange stronger along one bond direction than another~\cite{kk1973}. The orbitals themselves can also be excited. A propagating magnon can then transfer energy to orbital motion, altering both its dispersion and its damping~\cite{ko1997}. Orbital correlations have been invoked to explain precisely this combination of softening and damping in manganites~\cite{ref9}. A directional exchange path can thus influence both the energy of a magnon and the states into which it can scatter.

\fig{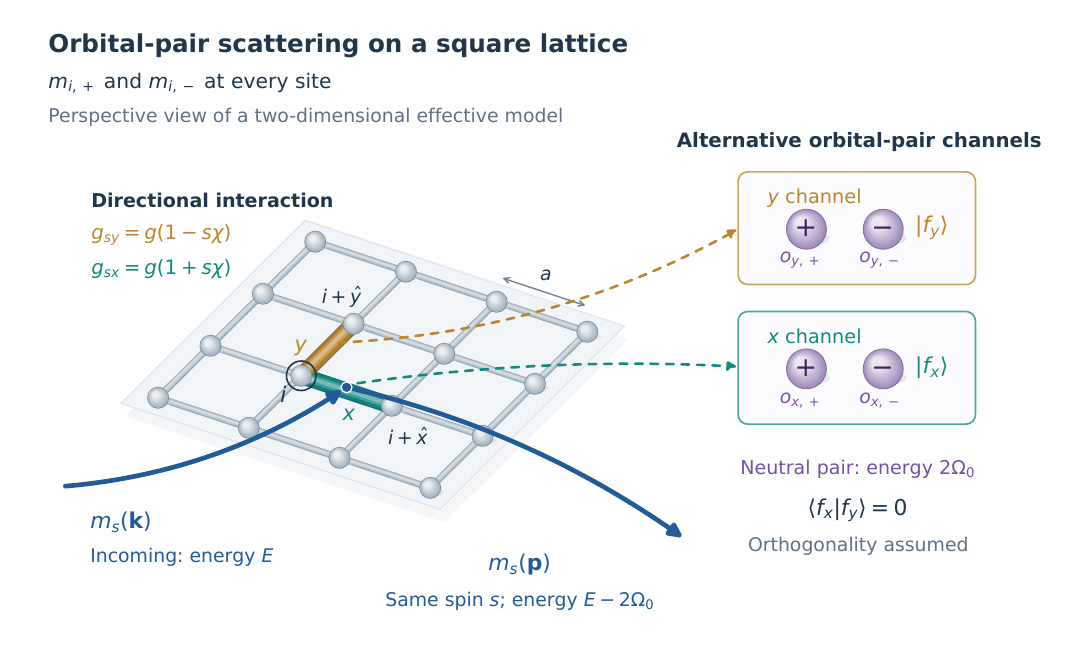}{0.87}{Emission of an orbital pair by a magnon on a square lattice of spacing $a$. The incoming blue arrow carries energy $E$ and spin $s$, and the outgoing arrow carries the same spin and energy $E-2\Omega_0$. The boxes show the two possible final states, a pair on an $x$ bond (teal) or a pair on a $y$ bond (orange). Each pair contains opposite orbital pseudospins. The bond couplings $g_{sx}$ and $g_{sy}$ set the emission amplitudes. Since $\langle f_x|f_y\rangle=0$, the two emission probabilities add.}

Direction matters particularly in compensated magnets, where the net magnetization vanishes. In an altermagnet, symmetry allows a spin splitting that changes sign between symmetry-related directions~\cite{ref1}, with a corresponding directional dependence predicted for magnons of opposite chirality~\cite{smejkal2023}. Orbital order can supply the inequivalent exchange paths~\cite{ref2}, and coupled spin--orbital models can support magnon and orbital branches that remain distinct in harmonic theory~\cite{ref3}. Previous studies have found unequal magnon lifetimes through spontaneous magnetic decay~\cite{ref4,ref5} and electronic Landau damping~\cite{ref6,ref7}. Spectroscopy can, in principle, distinguish an asymmetry in linewidth from a splitting in energy~\cite{ref8}. This distinction motivates a test in which the energy splitting is tuned to zero and the orbital contribution to each linewidth can be compared directly.

The square-lattice model in Fig.~\ref{fig:1} implements this test with a single elementary process. A magnon emits two orbital excitations and continues with the same spin. Along a given lattice direction, one spin branch couples more strongly to the orbital pair than the other. An independent exchange anisotropy can then be adjusted until the two interacting magnons have the same resonance energy. Throughout this paper, \emph{compensation} refers to this equality at a specified momentum. It is achieved by tuning a parameter. A symmetry that exchanged the two branches at that same momentum would also require their linewidths to agree.

With the energies matched, the linewidths reveal how strongly each spin branch scatters. The model also predicts where to look for the difference. A quarter turn exchanges the preferred bond directions of the two branches and hence interchanges their lifetimes. If the two branches are injected with equal intensity, the longer-lived branch contributes more of the signal that reaches a distant detector within the incident energy window.

To look for this effect in a material, one must first identify which excitations can receive the magnon's energy and momentum. Spin--orbit-coupled $d^4$ magnets offer a useful comparison because they support collective oscillations of the ordered moment~\cite{khaliullin2013}. In Ca$_2$RuO$_4$, neutron and Raman measurements identify an amplitude mode and its coupling to magnetic excitations~\cite{ref10,souliou2017}, and resonant inelastic x-ray scattering (RIXS) resolves the spin--orbital character of the low-energy modes~\cite{ref11,ref12}. The comparison developed below concerns the decay of this amplitude mode into two transverse magnetic modes. It illustrates how to identify a decay threshold from the final states. The Supplemental Material~\cite{supplement} gives the orbital-pair model comparisons in Sec.~\smsec{5} and the material response in Sec.~\smsec{13}.

\section{Spin-conserving orbital-pair scattering}

To isolate the orbital process, consider a square lattice of spacing $a$ with two magnon branches, labeled $s=+1$ and $-1$ according to their spin. Each bond also supports orbital excitations with two opposite internal labels, called orbital pseudospins. In Fig.~\ref{fig:1}, the blue arrows follow a single magnon through a scattering event. It leaves a pair of orbital excitations on either the teal $x$ bond or the orange $y$ bond. The two members of the pair have opposite pseudospins, and the outgoing magnon retains the spin it had on arrival.

The interaction measures how much the magnon wave varies across a bond. If $m_{i,s}$ annihilates a magnon on site $i$, then $D_{i\mu s}=m_{i+\hat\mu,s}-m_{i,s}$ is the difference between neighboring amplitudes along $\mu=x,y$. The product $D_{i\mu s}^{\dagger}D_{i\mu s}$ scatters a magnon within branch $s$ and couples to orbital terms that create or remove a pair or scatter from an occupied bond. A spatially uniform magnon has zero bond difference. For a nonuniform wave, the pair-emission process shown in the figure is
\begin{equation}
 m_s(\mathbf k)\longrightarrow m_s(\mathbf p)+o_+(\mathbf q)+o_-(\mathbf k-\mathbf p-\mathbf q).
\label{eq:process}
\end{equation}
Here $\mathbf k$ and $\mathbf p$ are the incoming and outgoing magnon momenta. The orbital pair carries their difference, up to a reciprocal lattice vector. For orbital modes with a flat dispersion, each excitation costs the same energy $\Omega_0$ at every momentum. The magnon therefore loses $2\Omega_0$ in one emission event. Since the outgoing magnon must still carry at least its gap energy $\Delta$, the minimum energy of the three excitations together is $E_{\rm th}=\Delta+2\Omega_0$. Above this threshold, many combinations of outgoing momenta are possible. These states form the emission continuum.

\fig{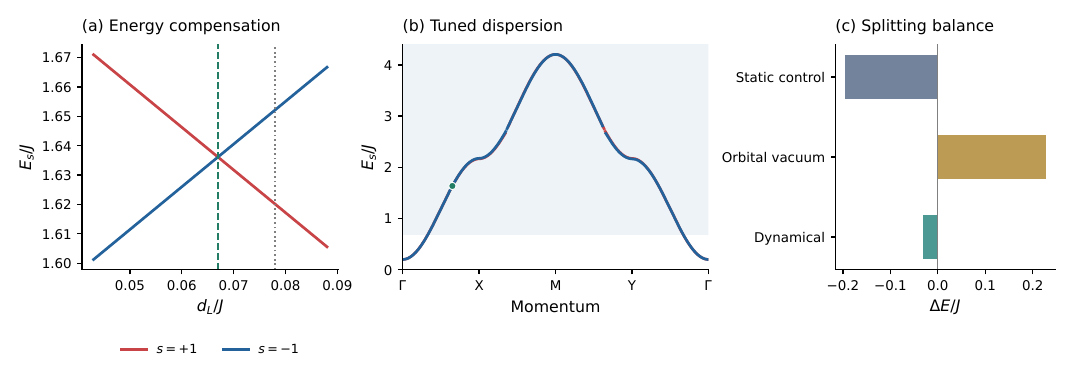}{0.90}{Tuning two magnon branches to the same resonance energy. (a) Red and blue curves give $E_+$ and $E_-$ at $\mathbf k_*$. The teal dashed line marks their crossing. The gray dotted line marks $d_L=2g\chi$, where the static splitting vanishes. (b) Dispersions at the crossing value of $d_L$, along $\Gamma=(0,0)$, $X=(\pi,0)$, $M=(\pi,\pi)$, $Y=(0,\pi)$, and back to $\Gamma$. The green point is the selected resonance. Shading marks energies above the emission threshold. Small breaks identify momenta that fail the narrow-resonance test. (c) Contributions to $E_+-E_-$ at the green point. The three signed shifts sum to zero.}
\fig{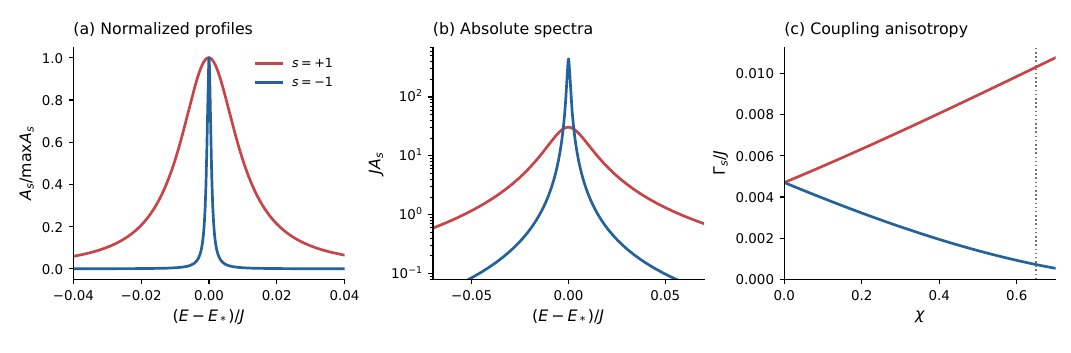}{0.90}{Two linewidths at the common resonance energy $E_*$. (a) Spectra for the $+$ (red) and $-$ (blue) branches, each divided by its peak height to expose the width difference. (b) Absolute spectra on a logarithmic scale, showing the height and tails of each peak. (c) Half-widths as the directional coupling $\chi$ is varied and the energies are retuned to equality. The dotted line marks $\chi=0.65$, used in (a) and (b).}

Whether a magnon can reach the continuum is an energy question. How readily it enters the continuum also depends on the bond it excites. The couplings are $g_{sx}=g(1+s\chi)$ and $g_{sy}=g(1-s\chi)$, where $g$ fixes the overall strength and $\chi$ its directional imbalance. For positive $\chi$, the $+$ branch couples more strongly to an $x$ bond, and the $-$ branch to a $y$ bond. Even in the orbital vacuum, the interaction contributes a static shift to the magnon dispersion. Including it gives $\epsilon_s(\mathbf k)=\Delta+Jh(\mathbf k)+sd f(\mathbf k)$, where $h=2-\cos k_x-\cos k_y$, $f=\cos k_x-\cos k_y$, and $d=d_L-2g\chi$. Here $J$ sets the energy scale, momenta are in units of $a^{-1}$, and the adjustable parameter $d_L$ controls the static exchange anisotropy.

The Hamiltonian in Supplemental Sec.~\smsec{1} has a stable vacuum for the parameters used here. Starting from this vacuum, the simplest calculation that includes emission retains states with zero or one orbital pair. Within this space a magnon can emit a pair, scatter repeatedly from the occupied bond, and reabsorb the pair. This is the one-pair calculation used below.

The location of the pair matters in this calculation. In Fig.~\ref{fig:1}, a pair on an $x$ bond and a pair on a $y$ bond are distinguishable final states, so their direct emission probabilities add. If several bonds excite the same pair on a shared site, their amplitudes can interfere. Supplemental Sec.~\smsec{9} defines this second geometry, and Sec.~\smsec{5} compares its linewidths with those for independent bonds. Starting with independent bonds allows the effect of directional coupling to be seen directly in the magnon spectrum.

\section{Compensated resonances and spectral contrast}

The quantity that connects the bond interaction to a spectral peak is the retarded self-energy $\Sigma_s$. Its real part shifts the magnon energy, and its imaginary part describes the loss of intensity into the available scattering states. Both follow from the same orbital-pair calculation (Supplemental Sec.~\smsec{2}). The spectral function, which gives the distribution of magnon weight over energy at a chosen momentum, is
\begin{equation}
 A_s(\mathbf k,E)=-\frac{1}{\pi}\Ima
 \frac{1}{E+i0^+-\epsilon_s(\mathbf k)-\Sigma_s(\mathbf k,E+i0^+)}.
\label{eq:spectral}
\end{equation}
For a well-resolved peak, its real-axis resonance energy $E_s$ and local half-width $\Gamma_s$ follow from
\begin{equation}E_s=\epsilon_s+\Rea\Sigma_s(\mathbf k,E_s),\label{eq:energy}\end{equation}
\begin{equation}\Gamma_s=-Z_s\Ima\Sigma_s,\qquad Z_s=[1-\partial_E\Rea\Sigma_s]^{-1}.\label{eq:width}\end{equation}
Equation~\eqref{eq:energy} locates the resonance by adding the interaction-induced shift to the magnon energy $\epsilon_s$. The factor $Z_s$ in Eq.~\eqref{eq:width} accounts for how rapidly the energy shift varies near that point. A smaller $\Gamma_s$ means a longer-lived magnon. Because a peak can be slightly asymmetric, its maximum need not fall precisely at $E_s$. The calculation uses $E_s$ to define compensation and independently measures the width from the spectral profile. A small numerical broadening $\eta$ enters the internal self-energy only. The external propagator in Eq.~\eqref{eq:spectral} retains $i0^+$.

Figure~\ref{fig:2}(a) shows how to reach compensation in practice. At zero temperature, take $\mathbf k_*=(0.65\pi,0)$, $\Delta=0.20J$, $\Omega_0=0.25J$, $g=0.06J$, and $\chi=0.65$. As $d_L$ increases, the red ($+$) resonance moves down and the blue ($-$) resonance moves up. Setting $d_L=2g\chi$ cancels the static anisotropy, but the two branches still have different resonance energies at the gray dotted line because orbital scattering shifts them by different amounts. The interacting energies coincide at the teal dashed line, where $d_L=0.06698J$ and $E_+=E_-=E_*=1.63612J$ for $\eta=0.003J$.

An energy crossing is useful for comparing lifetimes only if the decay channel is open there. Figure~\ref{fig:2}(b) checks this condition with $d_L$ fixed at the crossing. The green point at $\mathbf k_*$ lies above the threshold $E_{\rm th}=0.70J$ in the shaded continuum. Panel~(c) resolves the energy difference into three contributions. The positive orbital-vacuum shift is canceled by the negative static-control and dynamical terms. Their sum fixes the common energy. Each branch can still emit a pair, and its emission rate can now be read from the width of its peak.

Figure~\ref{fig:3}(a) puts the two peaks at the same height so that their widths can be compared directly. Near the common resonance, the red peak is much broader than the blue one. Restoring the absolute spectral weight in panel~(b) shows that the blue peak is taller because its weight is concentrated in a narrower energy interval. The calculated half-widths are $\Gamma_+=0.01030J$ and $\Gamma_-=0.0007203J$. Thus the negative-spin excitation lives about $14.3$ times as long as the positive-spin excitation at the same resonance energy. Widths measured at half maximum agree with Eq.~\eqref{eq:width} to within $0.0142\%$ (Supplemental Secs.~\smsec{3} and \smsec{17}).

Panel~(c) tests whether this difference comes from the directional coupling. At $\chi=0$, the two branches couple equally to the $x$ bond and have equal widths. As $\chi$ increases, the red peak broadens and the blue peak narrows, even though $d_L$ is retuned at each point to keep their energies equal. The dotted line marks the value used in panels~(a) and (b). Momentum-integration and grid checks appear in Supplemental Secs.~\smsec{10} and \smsec{11}, and an independent evaluation reproduces the reference ratio in Sec.~\smsec{16}.

\section{Compensation contour and a weak-coupling criterion}
\fig{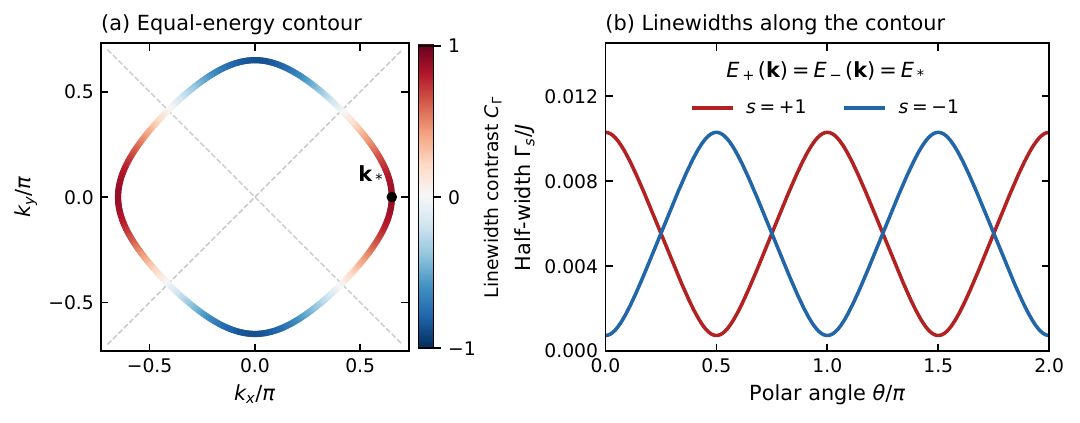}{0.94}{Which spin lives longer along the equal-energy contour? (a) Color gives the signed linewidth contrast $C_\Gamma$ on $h(\mathbf k)=h(\mathbf k_*)$. The $-$ branch lives longer on red arcs and the $+$ branch on blue arcs. The black dot marks $\mathbf k_*$ and dashed lines mark the diagonals. (b) Individual half-widths along the same contour, plotted against the polar angle of $\mathbf k$. The curves cross on the diagonals and exchange under a quarter turn, with $E_+=E_-=E_*$ throughout.}

At $\mathbf k_*$, the longer-lived branch is the one that couples weakly to the $x$ bond. Rotating the wave vector by $90^\circ$ exchanges the roles of the $x$ and $y$ bonds and, with them, the roles of the two spin branches. If $R$ denotes this rotation, symmetry gives
\begin{equation}
 E_+(R\mathbf k)=E_-(\mathbf k),\qquad
 \Gamma_+(R\mathbf k)=\Gamma_-(\mathbf k).
\label{eq:rotation}
\end{equation}
Consequently the $+$ branch at $R\mathbf k_*$ has the energy and lifetime of the $-$ branch at $\mathbf k_*$. This symmetry relates two different momenta. At a momentum where a symmetry exchanges the branches while leaving the momentum fixed, their entire spectral responses must coincide.

Rotation accounts for the exchange of lifetimes at two points. To find where else the energies coincide, consider the momentum dependence of the one-pair self-energy. For flat orbital modes it contains only the two lattice functions $h$ and $f$ introduced above,
\begin{equation}
 \Sigma_s(\mathbf k,z)=2[T_x(z)+T_y(z)]h(\mathbf k)
 -2s[T_x(z)-T_y(z)]f(\mathbf k),
\label{eq:separable}
\end{equation}
where $T_x$ and $T_y$ describe repeated scattering from an occupied bond (Supplemental Sec.~\smsec{6}). In the resonance equation, all terms that change sign between the two spins multiply the same function $f(\mathbf k)$. Tuning their real coefficient to zero at $E_*$ and at a point with $f(\mathbf k_*)\ne0$ removes that spin dependence at $E_*$. The remaining momentum dependence enters through $h$, so the same energy solves both branch equations everywhere on the contour
\begin{equation}h(\mathbf k)=h(\mathbf k_*).\label{eq:contour}\end{equation}
Figure~\ref{fig:4}(a) traces this contour through the black point $\mathbf k_*$. The color records the signed linewidth contrast $C_\Gamma=(\Gamma_+-\Gamma_-)/(\Gamma_++\Gamma_-)$. Along the red arcs the $-$ branch lives longer. Following the contour toward the $k_y$ axis leads into a blue arc, where the $+$ branch lives longer. Panel~(b) shows the two widths approach each other, cross on a diagonal, and exchange values after a quarter turn. On the diagonal each half-width is $0.00553J$ (Supplemental Sec.~\smsec{16}). The equality of energies along the entire curve follows from the particular momentum dependence in Eq.~\eqref{eq:separable}. The exchange under rotation follows from the lattice symmetry and is a separate prediction.

The size of the axial contrast has a simple origin at weak coupling. A wave traveling along $x$ has the same phase on neighboring sites along $y$, so its incoming $y$-bond difference vanishes. Only the $x$ bond contributes to emission. Inside a regular part of the continuum, the residual anisotropy required for compensation is $d=O(g^2/J)$. To leading order the two spins therefore have the same available final-state energies, and their emission rates differ through the squared couplings $g^2(1+s\chi)^2$. This gives
\begin{equation}
 \lim_{g\to0^+}\frac{\Gamma_+}{\Gamma_-}
 =\left(\frac{1+\chi}{1-\chi}\right)^2.
\label{eq:weak}
\end{equation}
For $\chi=0.65$, Eq.~\eqref{eq:weak} gives a limiting ratio of $22.22$. The limit is taken after the momentum spectrum becomes continuous and the numerical broadening tends to zero. Scattering from an occupied pair first changes the result at order $g^3$, and a second simultaneous pair enters at order $g^4$ (Supplemental Sec.~\smsec{5}). These corrections explain why the weak-coupling value and the finite-coupling ratio in Fig.~\ref{fig:3} can differ. Both absolute widths vanish as $g^2$ as the coupling tends to zero. Even a large ratio can then correspond to very little loss over a finite distance. For calculations retaining all configurations up to a specified number of pairs, Supplemental Sec.~\smsec{15} bounds the error in the self-energy at fixed energy and finite $\eta$.

\section{Momentum dependence and continuum access}
\fig{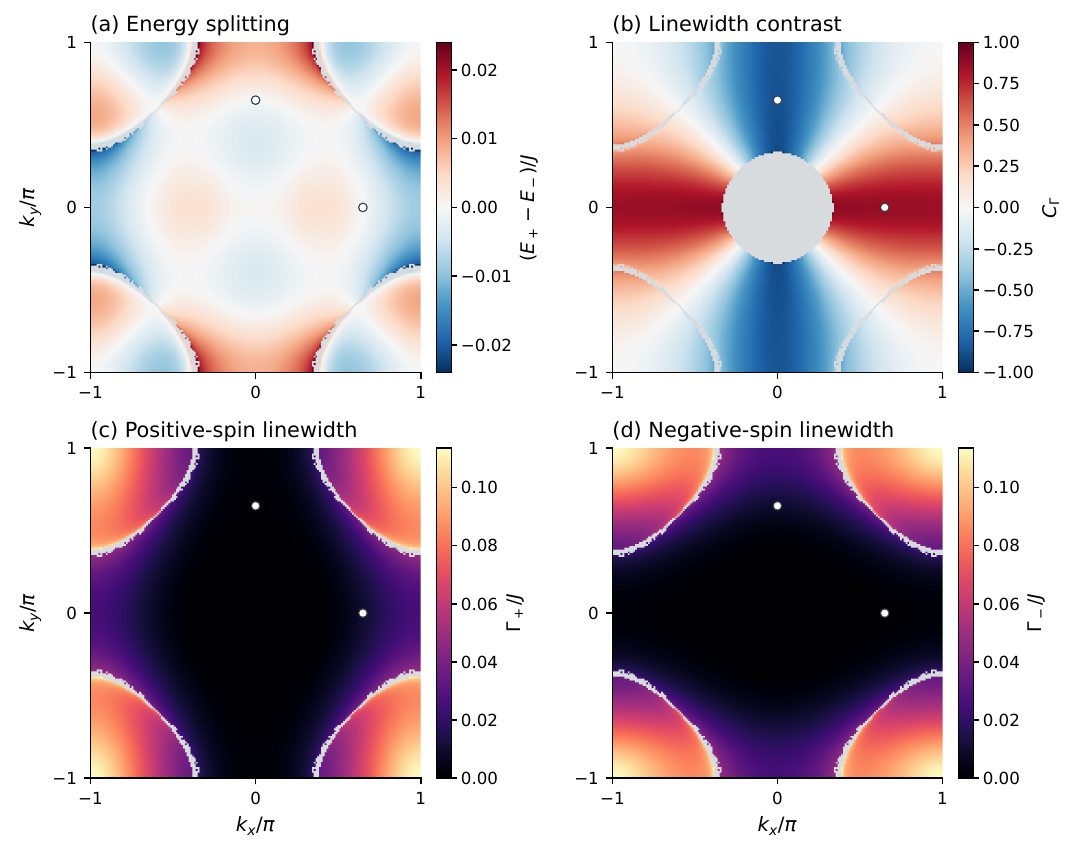}{0.89}{Energy splitting and linewidths across the Brillouin zone at the tuned value of $d_L$. (a) Signed energy difference $(E_+-E_-)/J$. (b) Signed linewidth contrast $C_\Gamma$. Red favors longer survival of the $-$ branch and blue favors the $+$ branch. (c,d) Half-widths $\Gamma_+/J$ and $\Gamma_-/J$ on a common scale. White markers identify $\mathbf k_*$ and its quarter-turned partner. Widths below threshold are set to zero. Gray marks points failing the resonance test and, in (b), points where both widths vanish. An AI-assisted plotting script renders the archived numerical arrays.}

Figure~\ref{fig:5} places the equal-energy result in the full Brillouin zone, the range of distinct wave vectors of the lattice. Panels~(a) and (b) show two quantities a spectroscopic measurement could distinguish. Panel~(a) gives the energy difference $E_+-E_-$, which vanishes at the two white markers. At the same markers, panel~(b) shows a sizable linewidth contrast with opposite signs. A measurement of the energy splitting alone would therefore miss both the lifetime difference and its reversal upon rotation.

Panels~(c) and (d) give the individual widths on the same color scale. A quarter turn maps the large widths of one branch onto those of the other, as required by Eq.~\eqref{eq:rotation}. Near the zone center, both maps are dark because neither branch has enough energy to emit a pair. That region is gray in panel~(b) because $C_\Gamma$ would involve division by zero. Elsewhere, gray identifies points at which a well-defined resonance could not be assigned under the criteria in Supplemental Sec.~\smsec{11}.

The onset of damping at the boundary of the central region follows from the bond interaction. Just above threshold, the outgoing magnon has momentum close to zero, so its wave varies little from one site to the next. The emission probability contains the bond factor $1-\cos p_\mu\simeq p_\mu^2/2$ for an outgoing magnon with momentum $\mathbf p$. Together with a quadratic band minimum in two dimensions, this gives a linear onset of absorption with energy above threshold (Supplemental Sec.~\smsec{6}). The same coupling produces the energy shift through the self-energy in Sec.~\smsec{2}. The numerical maps use flat orbital modes. If the orbitals disperse, the lowest final-state energy must be found separately for each total momentum, as described in Sec.~\smsec{12}.

Once emission is allowed, the quantitative question is how the linewidth ratio changes when more orbital pairs can participate. Retaining all configurations with up to two pairs gives a ratio of $15.5341$ after extrapolation to zero internal broadening. Supplemental Fig.~\ref{fig:conv}(a) tests the spatial grid at fixed broadening, and panel~(b) shows the extrapolation after grid refinement. Allowing up to eight pairs with a restriction on their separation gives a ratio near $15.62$ at $\eta=0.003J$. Panel~(c) varies the allowed separation in the three-pair sector to show the sensitivity to this restriction. The two-pair and higher-pair spaces are defined in Supplemental Secs.~\smsec{7} and \smsec{8}, with extraction, convergence procedures, and results in Secs.~\smsec{17}--\smsec{19}. The sharp-limit ratio for an unrestricted number of pairs remains undetermined.

The linewidth difference thus persists when the scattering calculation includes more pairs. Its observable consequence depends on how far magnons of each spin travel before losing energy.

\section{Attenuation of an energy-resolved magnon signal}
\fig{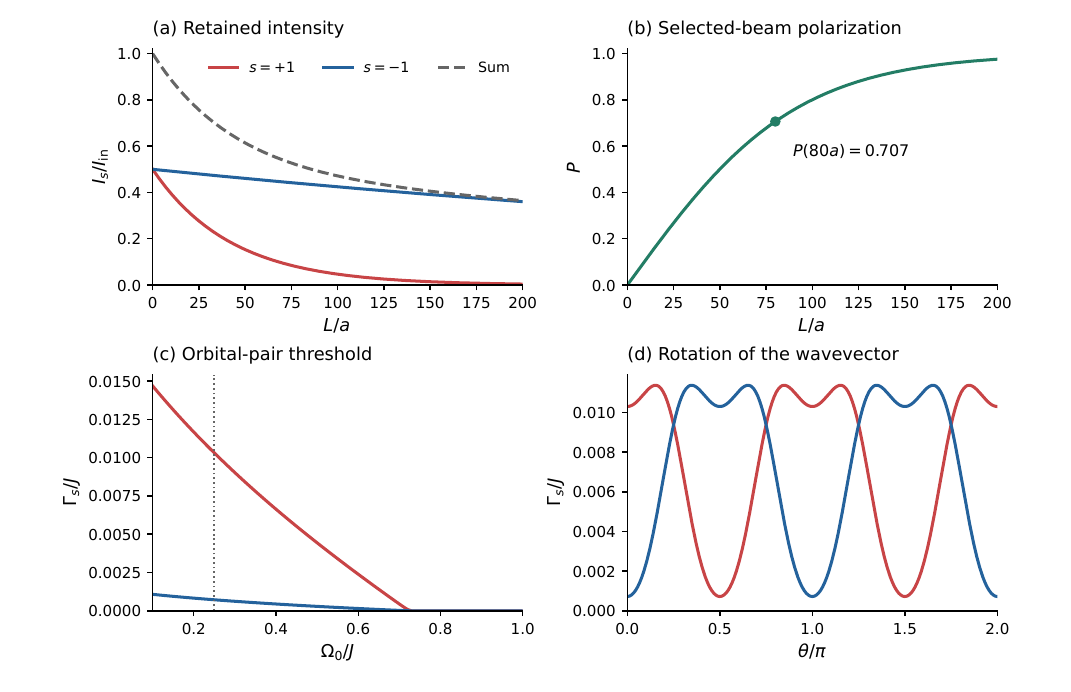}{0.91}{Spin selection during propagation. (a) Intensities remaining in the incident window for $s=+1$ (red) and $s=-1$ (blue), as fractions of the total incident intensity. The dashed curve is their sum. (b) Polarization of that surviving signal. The green dot marks $P(80a)=0.707$. (c) Half-widths as the orbital energy increases, with the magnon energies retuned to compensation. The dotted line marks the reference value $\Omega_0=0.25J$. Widths below threshold are shown as zero. (d) Half-widths as the wave vector rotates at fixed magnitude and fixed Hamiltonian parameters. A quarter turn interchanges the branches.}

Consider equal incident fluxes in the two spin branches at $\mathbf k_*$. A detector selects a narrow window around the incident energy and momentum. A magnon that emits a pair leaves this window because its energy falls by $2\Omega_0$. For ballistic propagation, with negligible scattering back into the detection window, the fraction of the initial total intensity that reaches a distance $L$ in branch $s$ is (Supplemental Sec.~\smsec{4})
\begin{equation}
 I_s(L)=\tfrac12e^{-L/\ell_s},\qquad
 \ell_s=\frac{\hbar|v_s|}{2\Gamma_s},
\label{eq:intensity}
\end{equation}
where $v_s$ is the group velocity along the propagation direction. The half-width $\Gamma_s$ corresponds to an intensity lifetime $\hbar/(2\Gamma_s)$. Multiplying that lifetime by the speed gives the attenuation length $\ell_s$.

Figure~\ref{fig:6}(a) shows the consequence of this relation. At the selected momentum the two group velocities are similar, but their calculated attenuation lengths are $\ell_+=42.28a$ and $\ell_-=610.57a$. The red intensity therefore falls rapidly with distance while much of the blue intensity survives. The dashed curve adds the two and gives the fraction of the incident signal still inside the detection window.

The same intensities also determine the spin composition of what survives. Defining positive polarization to mean an excess of the longer-lived $-$ branch gives
\begin{equation}
 P(L)=\frac{I_-(L)-I_+(L)}{I_-(L)+I_+(L)}.
\label{eq:polarization}
\end{equation}
Figure~\ref{fig:6}(b) gives $P=0.707$ at $L=80a$. At this distance panel~(a) gives a surviving fraction of $0.514$. Thus just over half of the incident intensity reaches the detector in the selected window, and about $85\%$ of that detected intensity belongs to the $-$ branch. With the convention $S^z=\hbar(N_+-N_-)$, the signed spin polarization is $-P$.

The detector sees a spin imbalance because it selects a particular part of the spectrum. Each scattered $+$ magnon still contributes its original spin at a lower energy. Including the scattered population restores the spin balance of the incident beam.

The last two panels of Fig.~\ref{fig:6} identify two ways to control this selection. In panel~(c), raising $\Omega_0$ makes a pair increasingly expensive to emit. Both widths decrease and eventually vanish when the threshold rises above the compensated resonance. In panel~(d), the orbital energy and all other Hamiltonian parameters are held fixed while the direction of travel rotates at fixed $|\mathbf k|$. The widths meet on the diagonals and exchange under a quarter turn, reversing which spin is favored. This angular scan follows a circle in momentum space. Figure~\ref{fig:4} follows the contour selected by the equal-energy condition.

\section{Discussion}
The attenuation lengths in Fig.~\ref{fig:6} depend on the absolute widths of both branches. Two microscopic effects influence those widths. First, a magnon can scatter again from a bond on which it has already created a pair. Supplemental Fig.~\ref{fig:order}(a) compares successive treatments of this scattering. Panel~(b) shows that repeated scattering reduces the width of the more strongly coupled branch most, bringing the finite-coupling ratio below its leading-order value. Second, different bonds can lead to the same orbital final state. In the shared-site model of Supplemental Sec.~\smsec{9}, both widths are smaller than for independent bonds [Supplemental Fig.~\ref{fig:shared}(b)]. The narrower branch undergoes the larger fractional reduction, so the ratio in panel~(a) rises from $14.3$ to $17.6$ at the reference coupling. Identifying the orbital states excited by each bond is therefore essential for predicting attenuation in a material.

\begin{figure*}[t]
\centering\includegraphics[width=0.94\textwidth]{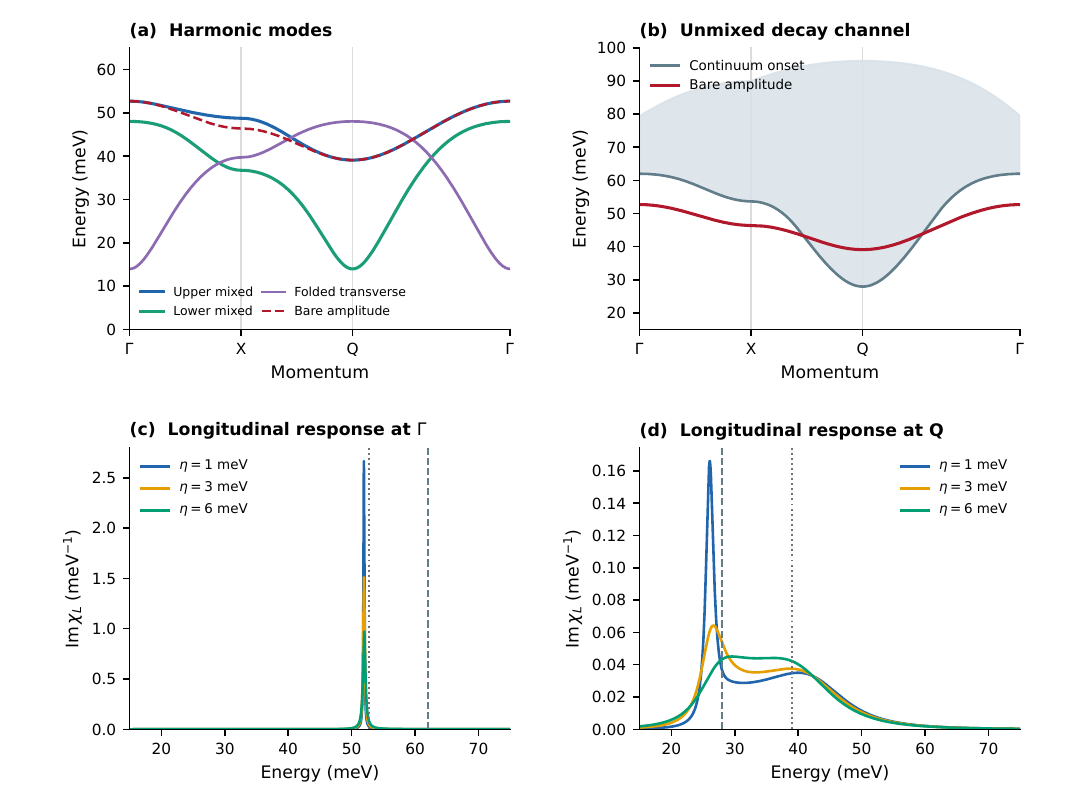}
\caption{Why the Ca$_2$RuO$_4$ amplitude mode has different spectra at the zone center and ordering wave vector, calculated with the model of Ref.~\cite{ref10}. (a) Harmonic modes along $\Gamma=(0,0)$, $X=(\pi,0)$, $Q=(\pi,\pi)$, and back to $\Gamma$. Blue and green denote the mixed modes, purple the folded transverse mode, and red dashed the bare amplitude mode. (b) Bare amplitude energy (red) and the range of two-transverse-mode energies (shaded), with the lower edge in blue-gray. (c,d) Longitudinal response $\operatorname{Im}\chi_L$ at $\Gamma$ and $Q$, respectively. Blue, orange, and green curves use internal broadenings $\eta=1$, 3, and 6~meV. Gray dotted lines mark the bare amplitude energies and blue-gray dashed lines mark the continuum thresholds. The vertical scales differ. Panels~(b)--(d) use unmixed transverse daughters.}
\label{fig:material}
\end{figure*}

Table~\ref{tab:candidates} lists materials in which different parts of this picture can be examined. In antiferromagnetic $\alpha$-Sr$_2$CrO$_4$, the $d_{xz}$ and $d_{yz}$ orbitals distinguish the two in-plane directions. Theory predicts an orbital mode that softens near a transition between staggered and entangled orbital order~\cite{ref14,ref15}. Such softening could lower the energy cost of emitting a pair, provided an experimentally accessible tuning path reaches this regime. In Sr$_2$VO$_4$, theory also predicts collective spin--orbital modes~\cite{ref16}. The magnetic state is less certain. One muon study reported static antiferromagnetism~\cite{ref17}, while another found fluctuating local fields down to 30~mK~\cite{yamauchi2015}. Establishing the ordered-state symmetry is therefore a prerequisite for a branch-resolved test in this compound.

\begin{table}[t]
\caption{Materials in which orbital excitations, magnetic symmetry, or the thresholds for magnetic decay can be examined.}
\label{tab:candidates}
\begin{ruledtabular}
\begin{tabular}{@{}p{0.25\columnwidth}p{0.70\columnwidth}@{}}
Material & Relevant ingredients \\
\hline
$\alpha$-Sr$_2$CrO$_4$ & Square-lattice antiferromagnetism; active orbitals; predicted orbital-mode softening~\cite{ref14,ref15}. \\[3pt]
Sr$_2$VO$_4$ & Active $d^1$ orbitals; predicted collective modes; low-temperature magnetic order unresolved~\cite{ref16,ref17,yamauchi2015}. \\[3pt]
Ca$_2$RuO$_4$ & Mixed spin--orbital excitations; amplitude-mode decay~\cite{ref10,ref11,ref12}. \\
\end{tabular}
\end{ruledtabular}
\end{table}

A test of orbital-pair emission in either candidate would begin with the measured magnetic and orbital dispersions. They must allow the incident magnon to supply the energy and momentum of two orbital excitations and an outgoing magnon. The symmetry of the ordered state must also permit unequal widths at the chosen wave vector. Under these conditions, branch-resolved spectra could track the widths as the resonance energies approach one another, then test for an exchange of the broader peak at a rotated wave vector. To interpret the measured intensity, one also needs the orbital character of the final modes, the coupling of the probe to each branch, and the populations of magnetic domains (Supplemental Sec.~\smsec{14}).

The first step of this material test, identifying the allowed final states, can already be illustrated in Ca$_2$RuO$_4$. Neutron scattering identifies a longitudinal amplitude mode coupled to transverse magnetic modes~\cite{ref10}. RIXS establishes the mixed spin--orbital character of the low-energy excitations~\cite{ref11,ref12}. Here a longitudinal oscillation changes the size of the ordered moment, while a transverse oscillation turns its direction. The material example in Fig.~\ref{fig:material} concerns the decay of one longitudinal oscillation into two transverse ones. Following Ref.~\cite{ref10}, the calculation starts from the pseudospin-one Hamiltonian
\begin{equation}
\begin{split}
H_{\mathrm{Ru}}={}&J_R\sum_{\langle ij\rangle}
(\mathbf S_i\cdot\mathbf S_j-\alpha S_i^zS_j^z)
+E_R\sum_i(S_i^z)^2\\
&+\varepsilon\sum_i(S_i^x)^2
+\sum_{\langle ij\rangle}A_{ij}(S_i^xS_j^y+S_i^yS_j^x).
\end{split}
\label{eq:Ru_model}
\end{equation}
where $\mathbf S_i$ acts on the local spin--orbital states. Exchange $J_R$ favors magnetic order, while the single-ion energy $E_R$ favors a local $S^z=0$ state. Their competition allows the magnitude of the ordered moment to fluctuate. The anisotropy $\alpha$ favors in-plane order, and $\varepsilon$ selects a direction in that plane. The bond term $A_{ij}$ changes sign between the two lattice directions and mixes longitudinal and transverse motion. Supplemental Sec.~\smsec{13} gives the parameters and response functions.

Figure~\ref{fig:material}(a) first identifies the modes whose energies enter the comparison. The red dashed curve is the amplitude mode before mixing. The blue and green curves follow the upper and lower mixed modes. Their separation near $X$ reflects the bond-dependent mixing in Eq.~\eqref{eq:Ru_model}, which vanishes at the zone center $\Gamma$ and at the ordering wave vector $Q=(\pi,\pi)$. The purple curve is the transverse mode shifted by $Q$, as expected when antiferromagnetic order folds the Brillouin zone.

To determine whether the amplitude mode can decay, panel~(b) compares its bare energy with the energies of two transverse modes carrying the required total momentum. Panels~(b)--(d) use unmixed transverse modes and the corresponding decay vertex. For an incident momentum $\mathbf q$, a daughter at $\mathbf k$ must be accompanied by one at $\mathbf q+\mathbf Q-\mathbf k$. The lower edge of the shaded region is the least energy such a pair can carry. At $\Gamma$, the amplitude energy is $52.71$~meV and the pair costs at least $62.01$~meV. At $Q$, both daughters can occupy low-energy states near the ordering wave vector. The pair threshold drops to $27.94$~meV, below the bare amplitude energy of $39.09$~meV.

This reversal in the order of the two energies explains the spectra in panels~(c) and (d). At $\Gamma$ [panel~(c)], the amplitude mode lies below the continuum and produces a narrow peak. Even below threshold, coupling to intermediate two-mode states shifts the resonance downward, giving a pole at $51.923$~meV in the sharp-mode limit. Increasing the prescribed broadening of the transverse modes, $\eta=1$, 3, or 6~meV, extends their spectral tails below the nominal edge and broadens the plotted peak. A separate external broadening of $0.05$~meV is used to display the response.

At $Q$ [panel~(d)], a transverse pair can carry the bare amplitude energy, and longitudinal spectral weight spreads into the continuum. The same coupling also shifts some weight into a feature below the edge. As the internal broadening decreases, this feature sharpens toward a pole at $25.993$~meV. The curves in (c) and (d) use the same three input broadenings, with different vertical scales for the two responses. These broadenings test how well the spectral features can be resolved. The pole energies and checks of their convergence with momentum-grid size are given in Supplemental Secs.~\smsec{13} and \smsec{16}.

The comparison with Ca$_2$RuO$_4$ shows what a threshold analysis can establish. The measured character of the modes identifies a magnetic two-mode decay channel, and momentum conservation explains why that channel produces different spectra at $\Gamma$ and $Q$. Establishing orbital-pair emission in a candidate material requires the corresponding identification of two excitations with orbital weight and a surviving magnon of the incident spin. Only then can a branch-dependent linewidth be assigned to the mechanism in Fig.~\ref{fig:1}.

\section{Conclusions}
In the square-lattice model, a magnon leaves its initial energy and momentum state by emitting two orbital excitations without changing its spin. Tuning the exchange anisotropy brings the opposite-spin resonances to the same energy, but their unequal couplings to an orbital pair still give them different lifetimes. The weak-coupling result identifies the squared bond amplitudes that determine the contrast. At the representative finite coupling, the linewidth ratio is $14.30$ in the one-pair calculation and extrapolates to $15.53$ when all states with up to two pairs are retained.

A direct spectroscopic signature would combine two observations. At a chosen momentum, the two resonance energies would meet while their widths remained distinct. At the quarter-turned momentum, the branch with the narrower line would switch. For flat orbital modes in the one-pair model, the entire equal-energy contour exhibits this directional pattern. In propagation, the narrower branch survives over a longer distance and comes to dominate the energy-selected signal, with the scattered magnons carrying the balance of the spin.

Connecting these signatures to a material requires knowing which excitations receive the energy. In Ca$_2$RuO$_4$ the relevant final states in the comparison are two transverse magnetic modes, and their threshold accounts for the contrast between $\Gamma$ and $Q$. For orbital-pair emission, the decisive material test would combine orbital spectroscopy of the proposed final states with a spin-resolved measurement of the two magnon widths at compensation.

\begin{acknowledgments}
J. S. acknowledges support from Uppsala University.
\end{acknowledgments}

\clearpage
\onecolumngrid
\begin{center}
\hypertarget{supplement.start}{}%
{\large\bfseries Supplemental Material\par}
\medskip
{\bfseries Equal-energy magnons with unequal lifetimes:\\
Orbital-pair scattering and a comparison with amplitude-mode decay in Ca$_2$RuO$_4$\par}
\end{center}
\twocolumngrid

\setcounter{section}{0}
\setcounter{equation}{0}
\setcounter{figure}{0}
\renewcommand{\thesection}{S\arabic{section}}
\renewcommand{\theequation}{S\arabic{equation}}
\renewcommand{\thefigure}{S\arabic{figure}}
\newcommand{\ssec}[2]{\section{#2}\label{sec:#1}}
\newcommand{\re}{\operatorname{Re}}
\newcommand{\im}{\operatorname{Im}}

\ssec{1}{Hamiltonian and stability}
A magnon couples to the orbital modes on bond $(i,\mu)$ through
$D_{i\mu s}=m_{i+\hat\mu,s}-m_{i,s}$ and
$C_{i\mu}=o_{i\mu,+}+o^\dagger_{i\mu,-}$. The Hamiltonian is
\begin{align}
H={}&\sum_{\mathbf k s}\epsilon^b_s m^\dagger_{\mathbf k s}m_{\mathbf k s}
 +\Omega_0\sum_{i\mu\nu}n_{i\mu\nu}\nonumber\\
 &+\sum_{i\mu s}g_{s\mu}D^\dagger_{i\mu s}D_{i\mu s}
 C^\dagger_{i\mu}C_{i\mu}.
\end{align}
Here $\epsilon_s^b=\Delta+(J-2g)h+sd_L f$,
$h=2-\cos k_x-\cos k_y$, $f=\cos k_x-\cos k_y$, and
$(g_{sx},g_{sy})=g(1+s\chi,1-s\chi)$. Momenta are in units of $a^{-1}$.
For $g\ge0$ and $|\chi|<1$ the interaction is positive semidefinite.
Since $|f|\le h$, the vacuum is stable for $J-2g>|d_L|$ and
$\Delta,\Omega_0>0$. The vacuum term $\langle C^\dagger C\rangle=1$
gives $\epsilon_s=\Delta+Jh+sdf$, with $d=d_L-2g\chi$.
The interaction conserves each magnon spin population and creates or
removes orbital excitations in opposite-pseudospin pairs.

\ssec{2}{Orbital-pair self-energy}
Write $F_\mu(\mathbf p)=1-\cos p_\mu$. The emission vertex
$g_{s\mu}(e^{ik_\mu}-1)(e^{-ip_\mu}-1)$ gives a squared amplitude
$4g_{s\mu}^2F_\mu(\mathbf k)F_\mu(\mathbf p)$. For flat, independent
bond modes the normalized sum over the internal orbital momentum is unity. Define
\begin{align}
 I_{s\mu}(z)&=\int_{\rm BZ}\frac{d^2p}{(2\pi)^2}
 \frac{F_\mu(\mathbf p)}{z-\epsilon_s(\mathbf p)-2\Omega_0},\\
 \Sigma_s(\mathbf k,z)&=4\sum_\mu F_\mu(\mathbf k)
 \frac{g_{s\mu}^2 I_{s\mu}(z)}{1-4g_{s\mu}I_{s\mu}(z)} .
 \label{eq:kernel}
\end{align}
For a pair fixed on one bond, the magnon sees the potential
$2g_{s\mu}|b\rangle\langle b|$, with $|b\rangle=|i+\hat\mu\rangle-|i\rangle$.
The free bond propagator is $2I_{s\mu}$, so its rank-one resummation
gives Eq.~\eqref{eq:kernel}, exact within the zero- and one-pair space~\cite{ref13}.

\ssec{3}{Compensated one-pair resonance}
At $T=0$ take $(\Delta,\Omega_0,g)/J=(0.20,0.25,0.06)$,
$\chi=0.65$ and $\mathbf k_*=(0.65\pi,0)$. At $\eta=0.003J$,
compensation gives $d_L=0.06697652J$ and $E_*=1.63611994J$.
The local half-widths are $\Gamma_+/J=0.01030233$ and
$\Gamma_-/J=0.000720278$, giving $\Gamma_+/\Gamma_-=14.30327$.
The spectral half-widths agree within $0.0142\%$.

\ssec{4}{Attenuation of the selected signal}
At fixed Hamiltonian parameters, $v_s=(a/\hbar)\partial_{k_x}E_s$ and
$\ell_s=\hbar|v_s|/(2\Gamma_s)$. For equal incident fluxes and negligible
reinjection into the detection window, $I_s(x)=e^{-x/\ell_s}/2$.
The lengths $\ell_+=42.28a$ and $\ell_-=610.57a$ give a surviving
fraction $I_-+I_+=0.514$ and polarization $(I_--I_+)/(I_-+I_+)=0.707$
at $x=80a$. Scattered magnons retain their spin.

\begin{figure*}[t]
\centering\includegraphics[width=0.79\textwidth]{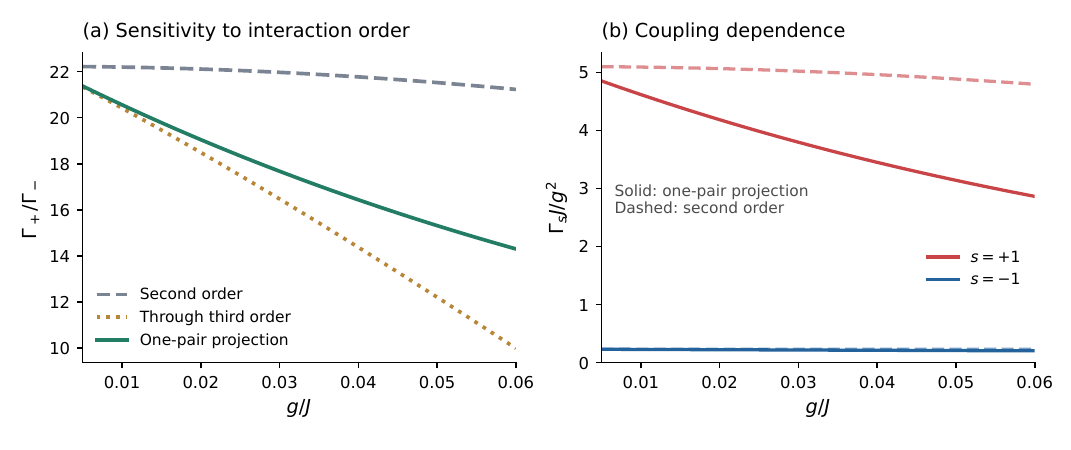}
\caption{(a) Linewidth ratio at compensated energy
as the interaction order and coupling vary. (b) Branch half-widths in units of
$g^2/J$ for the second-order and one-pair calculations.}\label{fig:order}
\end{figure*}

\begin{figure*}[t]
\centering\includegraphics[width=0.79\textwidth]{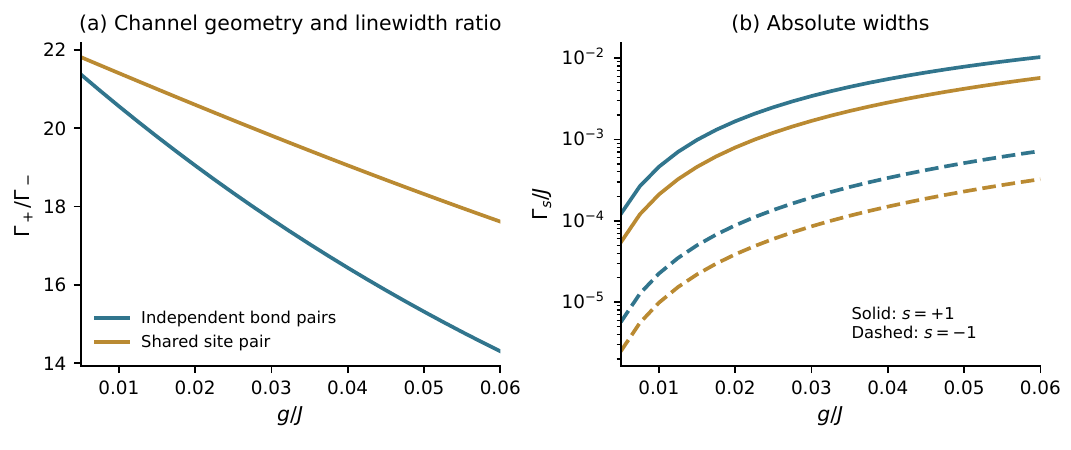}
\caption{Independent bond pairs and a shared site pair
at $\mathbf k_*$. (a) Compensated linewidth ratio. (b) Absolute half-widths
for the broad ($+$) and narrow ($-$) branches.}\label{fig:shared}
\end{figure*}

\ssec{5}{Interaction order and channel geometry}
Along $k_y=0$ only the incident $x$ bond contributes. Inside a regular
continuum, compensation gives $d=O(g^2/J)$ and
$\Gamma_s\propto g^2(1+s\chi)^2$ to leading order. Taking the continuum
and retarded limits before $g\to0$ gives a ratio of $[(1+\chi)/(1-\chi)]^2=22.2245$.
Scattering from an occupied pair first enters at order $g^3$ and a
second simultaneous pair at order $g^4$. At $g=0.06J$, retuning at each
order gives ratios of $21.23$ (second order), $9.99$ (through third
order), and $14.30$ (one pair). The occupied bond suppresses the broad
branch most strongly (Fig.~\ref{fig:order}). Sharing the final orbital
state between bonds reduces both widths, with a larger fractional
reduction for the narrow branch (Fig.~\ref{fig:shared}). Adding a second
pair gives a ratio of $15.54$ at $\eta=0.003J$. Sections~\ref{sec:7}
and \ref{sec:8} define the multipair spaces.


\ssec{6}{Emission threshold and compensation contour}
The continuum starts at $E_{\rm th}=\Delta+2\Omega_0$. Near its
quadratic minimum, $F_\mu(\mathbf p)\simeq p_\mu^2/2$ gives a linear
onset of absorption in two dimensions. Write
$T_\mu=g_{+\mu}^2 I_{+\mu}/(1-4g_{+\mu}I_{+\mu})$.
Quarter-turn symmetry exchanges $x,y$ and $+,-$, so
$\Sigma_s=2(T_x+T_y)h-2s(T_x-T_y)f$. At $f(\mathbf k_*)\ne0$,
compensation sets $d=2\re(T_x-T_y)$ at $E_*$. The same root therefore
holds along $h(\mathbf k)=h(\mathbf k_*)$ in this one-pair model.

\ssec{7}{Complete two-pair projection}
Let $P_j$ project onto $j$ orbital pairs, $H_{ij}=P_iHP_j$, and
$u=H_{10}|\mathbf k,s;0\rangle$. Eliminating $P_2$ gives
\begin{align}
 K_2&=H_{12}(z-H_{22})^{-1}H_{21}, \nonumber\\
 \Sigma_s^{(\le2)}&=u^\dagger(z-H_{11}-K_2)^{-1}u.
 \label{eq:schur}
\end{align}
The two occupied bonds remain fixed inside $H_{22}$. Each distinct bond
contributes a rank-one potential. Creating a second pair on the same
bond doubles its creation matrix element. All separations are retained.

\ssec{8}{Higher-pair configurations}
Above two pairs, retain configurations whose maximum distance between
occupied bond centers is at most $D_n$. The magnon remains mobile. With eight pairs allowed, $D_3=4a$, $D_{4\ldots8}=\sqrt2a$,
$L=5120$ and $\eta=0.003J$, the ratio is $15.6240$. The free three-pair
threshold is $1.70J$, above $E_*$. Closed higher sectors still change the
widths through virtual scattering.

\ssec{9}{Shared orbital final states}
A site pair couples to four bonds $\nu=+x,-x,+y,-y$.
Set $B_\nu=|\hat\nu\rangle-|0\rangle$, $\ell_\nu=e^{i\mathbf k\cdot\hat\nu}-1$
and $C_{\nu\nu}=g_{s|\nu|}/2$ in this basis, with $C$ diagonal.
For $M=B^\dagger(z-H_m^{\rm ref}-2\Omega_0)^{-1}B$ and magnon
dispersion $\epsilon_s$, the shared-pair self-energy is
\begin{equation}
\Sigma_s^{\rm sh}=(C\ell)^\dagger M(1-2CM)^{-1}(C\ell).
\end{equation}
The common final state allows interference. At compensation,
$\Gamma_+/J=0.00571468$ and $\Gamma_-/J=0.000324447$, giving $17.6136$.

\ssec{10}{Momentum integration}
Set $A=J-sd$, $B=J+sd$ and $w=z-\Delta-2J-2\Omega_0+A\cos p_x$.
The continuous $p_y$ average of $(w+B\cos p_y)^{-1}$ is $1/S$,
where $S=\sqrt{w-B}\sqrt{w+B}$ has the retarded branch $S\sim w$.
The even-$N$ midpoint average is $(1-r^N)/[S(1+r^N)]$ with
$r=-B/(w+S)$. Summing the remaining $p_x$ values gives the full
$N^2$ mesh. The sharp-limit quadrature splits the continuous $p_x$
integral at $w=\pm B$.

\ssec{11}{One-pair grid and map checks}
At $\eta=0.003J$, doubling the $4096^2$ midpoint mesh along both axes
changes the width ratio by less than $9\times10^{-6}$ relatively.
The independent sharp-limit value is $14.29404$. On the $181^2$
external momentum map, both branches must satisfy
$|E_s-\epsilon_s-\re\Sigma_s|<10^{-6}J$, $0.75<Z_s<1.25$,
$|\partial_E\Sigma_s|<0.25$ and $0\le\Gamma_s<0.1E_s$.
Plotted widths are zero below threshold. Points failing a resonance
test or having two zero widths are masked in the contrast map.

\ssec{12}{Dispersive orbitals}
For $\Omega(\mathbf q)=\Omega_0+t_o h(\mathbf q)$ the free continuum
edge at $\mathbf k$ is the minimum of
$\epsilon_s(\mathbf p)+\Omega(\mathbf q)+\Omega(\mathbf k-\mathbf p-\mathbf q)$.
The daughters can separate, so scattering from their individual
occupations requires a three-particle resolvent. The numerical
orbital-pair results use $t_o=0$.

\ssec{13}{Amplitude-mode decay in \texorpdfstring{Ca$_2$RuO$_4$}{Ca2RuO4}}
Main-text Fig.~\ref{fig:material} uses the Hamiltonian in Eq.~\eqref{eq:Ru_model} and follows Jain \emph{et al.}~\cite{ref10}. The inputs
$(J_R,E_R,\varepsilon,|A_{ij}|)=(5.8,25,4,2.3)$ are in meV and
$\alpha=0.15$. Define $W=8J_R$, $\tau=W/E_R$,
$\gamma_{\mathbf q}=(\cos q_x+\cos q_y)/2$, $u_0=W(1+\tau^{-1})/2$ and
$\lambda=(1-\alpha)(\tau-1)/(\tau+1)$. The harmonic energies follow from
\begin{align}
 \omega_L^2&=W^2(1+\gamma_{\mathbf q}/\tau^2),
 &\omega_T^2&=a_{\mathbf q}b_{\mathbf q},\nonumber\\
 a_{\mathbf q}&=u_0(1+\gamma_{\mathbf q})+\varepsilon,
 &b_{\mathbf q}&=u_0(1-\lambda\gamma_{\mathbf q})+\varepsilon .
\end{align}
In panel (a) the mixed squared energies are the eigenvalues of the
symmetric matrix with diagonal $(\omega_L^2,\omega_T^2)$ and off-diagonal
$\sqrt{W^3b_{\mathbf q}}\,|A_{ij}|(\cos q_x-\cos q_y)/(2J_R\tau)$.
The folded mode is $\omega_T(\mathbf q+\mathbf Q)$, $\mathbf Q=(\pi,\pi)$.

Panels (b)--(d) use unmixed transverse daughters with
$\mathbf k'=\mathbf q+\mathbf Q-\mathbf k$ and
$\omega_{\mathbf k}=\omega_T(\mathbf k)$. On a periodic $4096^2$ grid,
\begin{align}
 V^2&=\frac{W^2}{4}(1-\tau^{-2})
 \left(\frac{\gamma_{\mathbf q}}{\tau}
       +\frac{\gamma_{\mathbf k}+\gamma_{\mathbf k'}}{2}\right)^2,\nonumber\\
 \Pi_L&=\frac1N\sum_{\mathbf k}
 \frac{V^2 b_{\mathbf k}b_{\mathbf k'}
 (1/\omega_{\mathbf k}+1/\omega_{\mathbf k'})}
 {(\omega_{\mathbf k}+\omega_{\mathbf k'})^2-(\omega+i\eta)^2},\nonumber\\
 \chi_L^{-1}&=2[\omega_L^2-(\omega+i\delta)^2]/W-\Pi_L .
\end{align}
The plotted $\im\chi_L$ uses internal broadenings $\eta=1,3,6$ meV and
external $\delta=0.05$ meV. At $\Gamma$, the bare amplitude energy
(52.71 meV) lies below the two-daughter threshold (62.01 meV). At
$\mathbf Q$, it lies above (39.09 versus 27.94 meV). In the sharp kernel
the below-edge poles are 51.923 and 25.993 meV, respectively.

\ssec{14}{Projection onto material modes}
Expand the microscopic Hamiltonian about its ordered state and transform
the interaction into its harmonic mode basis. For incident branch
$\lambda$ and final state $f$, the leading decay weight is
$\pi|\langle f|H_{\rm int}|\lambda\rangle|^2\delta(E_\lambda-E_f)$.
An orbital-pair assignment requires two daughters with orbital weight
and a third carrying the outgoing magnon spin character. A symmetry
exchanging the incident branches at fixed $\mathbf k$ enforces equal
linewidths. Comparison with experiment also requires the probe weights
$|\langle\lambda|O|0\rangle|^2$ and the domain populations.

\begin{figure*}[t]
\centering\includegraphics[width=0.84\textwidth]{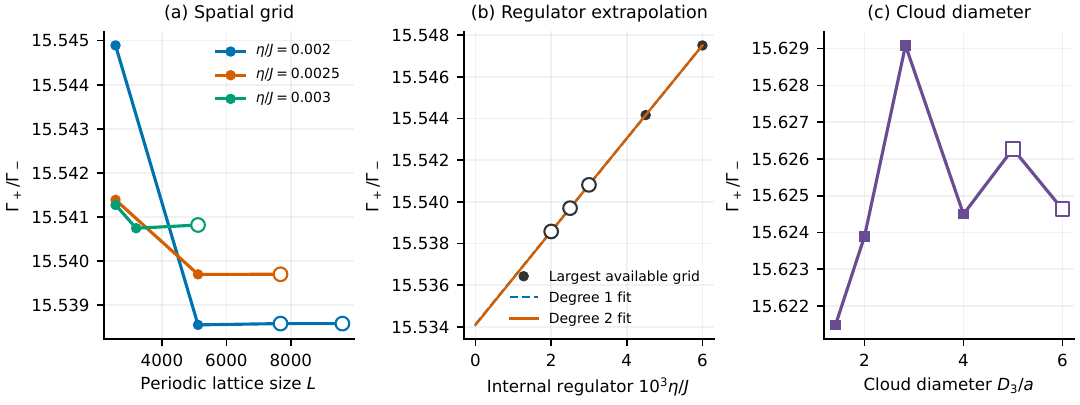}
\caption{Convergence of $\Gamma_+/\Gamma_-$. (a) Periodic grid size at fixed
$\eta$. (b) Regulator extrapolation in the complete two-pair space.
(c) Three-pair diameter at fixed higher-pair cutoffs.}\label{fig:conv}
\end{figure*}

\ssec{15}{Bound on omitted pairs}
For the complete pair-number cutoff $n$, positivity gives
$H\ge\Delta+2\Omega_0 N_{\rm pair}$ in the one-magnon sector.
For fixed spin write $g_\mu=g_{s\mu}$ and $g_m=\max_\mu g_\mu$.
The creation block from $j$ to $j+1$ pairs obeys
$\|H_{j+1,j}\|^2\le\beta_j^2=4(j+1)[jg_m^2+2(g_x^2+g_y^2)]$.
At $z=E+i\eta$, let $j_0$ be the first closed sector above one pair and
$D_j^2=(\Delta+2j\Omega_0-E)^2+\eta^2$. Orthogonality of successive
pair sectors bounds the tail by a factor
$\gamma_j=\beta_{j-1}/\sqrt{D_j^2+\beta_{j-1}^2}$ per step. The
resolvent identity then gives, for $n\ge j_0$,
\begin{equation}
 |\Sigma-\Sigma_{\le n}|\le
 \frac{\|u\|^2}{\eta^2}\,\beta_n\gamma_{n+1}
 \prod_{j=j_0}^{n}\gamma_j^2 .
\end{equation}
At $E=1.63548792J$, $\eta=0.003J$, $n=18$ suffices for a
$10^{-4}J$ self-energy error if every pair position is retained.
This fixed-energy bound leaves the unrestricted sharp-limit linewidth
ratio undetermined.

\ssec{16}{Independent numerical checks}
Direct summation reproduces the one-pair ratio $14.303267$. Across 361 contour points the resonance residual is below $2.4\times10^{-15}J$, and rotation exchanges the widths within $3.9\times10^{-17}J$. Both widths equal $0.00553428J$ on the diagonal. Independent four-bond inversion gives $17.61360$ for the shared channel. The Ca$_2$RuO$_4$ poles agree to the quoted precision on $512^2$ to $4096^2$ grids.

\ssec{17}{Methods for resonance and width extraction}
At $\mathbf k_*$ solve
$E_*-\epsilon_s(\mathbf k_*;d)-\re\Sigma_s(\mathbf k_*,E_*+i\eta;d)=0$
for both spins using a common $(E_*,d)$. At this root evaluate
\begin{equation}
 Z_s=[1-\partial_E\re\Sigma_s]^{-1},\qquad
 \Gamma_s=-Z_s\im\Sigma_s .
\end{equation}
The derivative is taken at fixed $d$. The regulator enters only the
internal self-energy. For the profile check in Sec.~\ref{sec:3}, locate
the maximum of $-\im[E+i0^+-\epsilon_s-\Sigma_s]^{-1}/\pi$ and the
two half-maximum crossings. Their separation gives the full width.

\ssec{18}{Methods for multipair convergence}
On a periodic $L\times L$ lattice with $L$ divisible by 40, solve
$(z-H_{11}-K_2)\psi=u$ for the $2L^2$ one-pair amplitudes. Fourier
transforms and local bond updates apply Eq.~\eqref{eq:schur}.
Differentiating the same resolvent gives the compensation Jacobian. The acceptance
criteria are a relative linear residual below $10^{-8}$ and a resonance
residual below $3\times10^{-10}J$ for each spin. Higher-pair configurations
are coupled through their boundary with the two-pair sector. Retune
compensation at each mesh, regulator and cutoff. Refine $L$ at fixed
$\eta$ before extrapolating $\eta\to0$. OpenAI Codex (GPT-6) assisted with numerical verification and documentation.

\ssec{19}{Convergence results}
At $\eta=0.002J$, the complete two-pair ratios on $L=5120$, 7680 and
9600 are $15.538545$, $15.538575$ and $15.538574$. At $\eta=0.003J$
and $L=5120$, the widths are $0.01142855J$ and $0.0007353894J$.
Linear and quadratic fits to the largest grids over windows within
$0.002\le\eta/J\le0.006$ give intercepts from $15.534087$ to $15.534118$.
The resulting sharp-limit estimate, $15.5341$, applies to the complete
two-pair space. Independent finite-volume Schur and Lanczos calculations
agree in self-energy within $1.89\times10^{-12}J$.

At $L=2560$ and $\eta=0.003J$, increasing $D_3/a$ from 4 to 5 to 6 gives
ratios $15.62450$, $15.62627$ and $15.62463$ with $D_{4\ldots8}=\sqrt2a$
(Fig.~\ref{fig:conv}). The nonmonotonic changes measure sensitivity to the
three-pair diameter. Larger four- through eight-pair clouds remain untested.

\end{document}